\documentclass[11pt,a4paper]{article}
\usepackage{jheppub}
\usepackage[T1]{fontenc}

\allowdisplaybreaks
\title{A renormalizable supersymmetric SO(10) model}
\author{Ying-Kang Chen}
\author{and Da-Xin Zhang}
\affiliation{School of Physics and State Key Laboratory of Nuclear Physics and Technology, \\
Peking University, Beijing 100871, China}

\emailAdd{ykchen@pku.edu.cn}
\emailAdd{dxzhang@pku.edu.cn}

\abstract{
A realistic grand unified model has never been constructed in the literature due to three major
difficulties: the seesaw mechanism
without spoiling gauge coupling unification,
the doublet-triplet splitting and the proton decay suppression.
We propose a renormalizable supersymmetric SO(10) model
with all these difficulties solved naturally by imposing
an extra discrete symmetry.}
\keywords{GUT, seesaw mechanism, doublet-triplet splitting, proton decay}
\begin{document}
\maketitle
\flushbottom
\pagenumbering{arabic}

\newcommand{\0}{SO(10)}
\newcommand{\5}{SU(5)}
\newcommand{\3}{$SU(3)_C$}
\newcommand{\2}{$SU(2)_L$}
\newcommand{\1}{$U(1)_Y$}
\newcommand{\Group}{$SU(3)_C\times
SU(2)_L\times U(1)_Y$}
\newcommand{\G}{$G_{321}$}
\newcommand{\group}{$SU(3)_C\times U(1)_{\textrm{em}}$}
\newcommand{\ii}{\textrm{i}}
\newcommand{\dd}{\textrm{d}}
\newcommand{\vev}[1]{\langle#1\rangle}
\newcommand{\ran}[1]{|#1\rangle}
\newcommand{\pair}{~$\bm{126}+\overline{\bm{126}}$~}
\newcommand{\del}{$\bm{126}$}
\newcommand{\delb}{$\overline{\bm{126}}$}
\newcommand{\sg}{\Sigma}
\newcommand{\sgb}{\overline{\Sigma}}
\newcommand{\dl}{\Delta}
\newcommand{\dlb}{\overline{\Delta}}
\newcommand{\be}{\begin{equation}}
\newcommand{\ee}{\end{equation}}
\newcommand{\bea}{\begin{eqnarray}}
\newcommand{\eea}{\end{eqnarray}}
\newcommand{\bref}[1]{(\ref{#1})}
\newcommand{\phd}[1]{\Phi^{(D)}_{#1}}
\newcommand{\pht}[1]{\Phi_{#1}}
\newcommand{\ba}{\begin{array}}
\newcommand{\ea}{\end{array}}

\section{Introduction}
The Supersymmetric (SUSY) Grand Unified Theory (GUT) is a very important direction
for the physics beyond the Standard Model (SM).
Among all the SUSY GUT models, the renormalizble SUSY SO(10) models\cite{clark,moha}
are very interesting
since they are very predictive\cite{msso10a,msso10a2,msso10b,msso10c,msso10d}.
All the fermions of every generation are contained in a spinor representation $\mathbf{16}$ ($\psi$),
which includes also a right-handed neutrino.
By coupling the $\psi$'s to the Higgs superfields in $\mathbf{10,\overline{126},120}$,
correct fermion masses and mixing can be generated while the right-handed neutrinos are given masses
by the Vacuum Expectation Value (VEV) from the SM singlet of $\overline{126}$
so that low energy neutrino oscillations are explained through the seesaw
mechanism\cite{seesawi1,seesawi2,seesawi3,seesawi4,seesawi5,seesawi6,seesawi7,seesawi8,seesawi9}.

There are three major difficulties in the SUSY SO(10) models to be solved.
First, there is a conflict between gauge coupling unification,
which requires a one-step breaking of the SO(10) group
into the SM gauge group at a GUT scale $\Lambda_{GUT}\sim 2\times 10^{16}{\rm GeV}$\cite{msso10b},
and the seesaw mechanism which requires the existence of a seesaw scale
$\Lambda_{seesaw}\sim 10^{-2} \Lambda_{GUT}$.
Such a lower intermediate scale not only breaks the unification,
but also lowers the masses of some color triplet masses which creates
more danger proton decay\cite{msso10b}.
Second,
proton decay is not suppressed to satisfy the present experimental limit.
The dominant mechanism for proton decay in SUSY GUTs is through the dimension-five operators mediated by the
color triplet  higgsinos coupling to the matter superfields.
These triplets have masses at $\Lambda_{GUT}$ which are not large enough.
Third,
there are a pair of weak doublets which are responsible
to the electro-weak  symmetry breaking in the Minimal SUSY SM (MSSM)
and must have weak scale masses $\sim 10^{2-3}{\rm GeV}$.
The large doublet-triplet splitting in masses,
which is more serious than the splitting in the  masses between  neutrinos and charged leptons
of order $10^{-6}\sim 10^{-9}$ to be explained through the seesaw mechanism,
needs to be naturally realized,
although the non-renormalization theorem of SUSY models can be used to stabilize this splitting when it is generated.
In the literature,
the Dimopoulos-Wilczek (DW)\cite{dw1,dw2,dw3,dw4,lee,dw6,dw7,dw8,dw9,dw10,dw11,dw12,dw13}
mechanism of missing VEV has been widely used in non-renormalizble models.
The doublet-triplet splitting in $\mathbf{10}$ is realized by coupling two different $\mathbf{10}$ to
$\mathbf{45}$ whose VEV is only in the (15,1,1) but not in the (1,1,3) direction,
where the numbers in the brackets are representations under the $SU(4)_C\times SU(2)_L\times SU(2)_R$ subgroup
of SO(10).
Since in $\mathbf{10}$ the doublets are in (1,2,2) and the triplets are in (6,1,1),
all the doublets are massless while their triplet partners are massive if the DW mechanism is applied.
There is also the Compliment to the DW (CDW) mechanism with another $\mathbf{45}$ whose VEV
is in the (1,1,3) direction only, coupling it with the second and a third massive $\mathbf{10}$
give masses to the second pair of doublets  while forbidding proton decay mediated by  the first $\mathbf{10}$.
However, neither the DW nor the  CDW mechanism can be applied directly in the renormalizable models in the presence of
$\mathbf{120}$ which contains also a pair of doublets in (15,2,2) whose existence invalidates
the DW mechanism.
Note that without $\mathbf{120}$ the supersymmetric SO(10) scenario
is not consistent with data \cite{bert}.

In the present work we are aiming at building a fully realistic model of renormalizable SUSY SO(10)
solving all the above difficulties, following the very recent progresses on solving
these difficulties by extending the Higgs sector.
To naturally realize
the model, an extra symmetry is enforced which is sufficient in constructing the
required superpotential.

\section{Review on the very recent progresses}
It has been realized in \cite{dlz} that to generate masses for the right-handed neutrinos,
it is a VEV of order $\Lambda_{seesaw}$ which is needed,
 instead of a symmetry breaking  scale above which
new particles emerge breaking gauge coupling unification.
$\Lambda_{seesaw}$ can be easily generated by introducing
two $\mathbf{\overline{126}}$s ($\overline{\Delta_{1,2}}$) and their conjugates $\mathbf{126}$s (${\Delta_{1,2}}$),
with only $\overline{\Delta_{1}}$ couples with the matter superfields.
Introducing the superpotential
\be
(m_{\Delta_{12}}+k \Phi) \overline{\Delta}_1 {\Delta}_2+
(m_{\Delta_{21}}+ \Phi) \overline{\Delta}_2 {\Delta}_1+Q\overline{\Delta}_2 {\Delta}_2,\label{W126}
\ee
where $\Phi$ is a $\mathbf{210}$ and $Q$ is a singlet, and we have suppressed
dimensionless couplings except $k$ which is the ratio of the first two trilinear couplings.
Maintaining SUSY requires both the D- and F-flatness conditions which are
\be
|\overline{v_{1R}}|^2+|\overline{v_{2R}}|^2=|v_{1R}|^2+|v_{2R}|^2\label{dterm}
\ee
and
  \bea 0&=&
  \left(\ba{cc}\overline{v}_{1R} & \overline{v}_{2R}\ea\right)
  \left(\ba{cc} 0 & m_{\text{$\Delta_{12} $}} +k\Phi_0\\m_{\text{$\Delta_{21} $}} +\Phi_0&Q \ea\right), \nonumber\\
  0&=&\left(\ba{cc} 0 & m_{\text{$\Delta_{12} $}} +k\Phi_0\\m_{\text{$\Delta_{21} $}} +\Phi_0&Q\ea\right)
  \left(\ba{c} {v}_{1R} \\ {v}_{2R} \ea \right),\label{fvr}
  \eea
respectively,
where $\Phi_0=\left[\Phi_1(1,1,1)\frac{1}{10\sqrt6}+\Phi_2(15,1,1)\frac{1}{10\sqrt2}+\Phi_3(15,1,3)\frac{1}{10}\right]$
is a combination of the three  VEVs of $\Phi$ responsible for SO(10) breaking,
and $v_{(1,2)R},\overline{v}_{(1,2)}$ are VEVs of $\Delta_{(1,2)},\overline{\Delta}_{(1,2)}$
responsible for $U(1)_{B-L}$ breaking.
For (\ref{fvr}) to have nontrivial solutions, the  determinant of the $2\times 2$ matrix needs be zero.
Physically,
since this matrix is also the mass matrix of the SM singlets whose VEVs ($\overline{v}_{iR}$ and ${v}_{iR}$)
break $U(1)_{B-L}$,
it must have one unique pair of Goldstone modes.
Then, choosing the (2,1) element of the $2\times 2$ matrix to be zero, the solutions to (\ref{fvr}) are
  \be
 \overline{v}_{1R}=\overline{v}_{2R}\frac{Q}{k m_{\Delta_{21}}-m_{\Delta_{12}} }, \label{v1r}
  \ee
and
  \be v_{2R}=0.\label{v2r}\ee
If we take
  \be Q\sim 10^{-2}\Lambda_{GUT},\label{Q}\ee
the seesaw scale VEV of the same order is generated for $\overline{v}_{1R}$ from (\ref{v1r}),
and $\overline{v}_{2R}\sim {v}_{1R}$ follows (\ref{dterm}).
Although this VEV is put in by hand, it may be linked with the (reduced) Planck scale as $Q\sim \frac{\Lambda_{GUT}^2}{M_{Planck}}$\cite{lz} by the Green-Schwarz mechanism\cite{green1,green2,green3,green4},
if we introduce a third pair of $\overline{126}-126$ which couple with a SO(10) singlet
which has an anomalous U(1) charge and gets a Planck scale VEV.

This small value $Q$ is also important in suppressing proton decay\cite{dlz2,lz}.
We also take (\ref{W126}) as an example.
There are three pairs of color triplets in each of $\overline{\Delta_{i}}+\Delta_{i}$,
and one more pair from $\Phi$.
The triplet mass matrix is
\be\label{masstoy}
\left(
\begin{array}{c|c}
0_{3\times 3}&(\Lambda_{GUT})_{3\times 4}\\
\hline
(\Lambda_{GUT})_{4\times 3}&C_{4\times 4}
\end{array}
\right),
\ee
where the triplets from $\Phi$ are put in the 4th column and the 4th row, and
\be
~C_{4\times 4}=\left(\begin{array}{cccc}
M_{\Phi^T} & v_{1R}&v_{1R}&0\\
\overline{v_{1R}}&0&0&Q\\
\overline{v_{1R}}&Q&0&0\\
0&0&Q&0\end{array}\right)
\ee
has three small eigenvalues of order $\overline{v_{1R}}\sim Q$ following (\ref{v1r}).
Then, after integrating out the triplets in $\Phi,\overline{\Delta}_2,{\Delta}_2$,
we can get the effective triplet mass matrix whose three eigenvalues are now all of the order
$\frac{\Lambda_{GUT}^2}{Q}$,
which are by a factor $O(100)$ larger than $\Lambda_{GUT}$ so that proton decay is suppressed.
This is an inverse analogue to the seesaw mechanism,
and it proposes a relation between the seesaw scale mass with the suppression of proton decay.
A practical model is presented for the Higgs sector with $\mathbf{10,\overline{126},{126},210}$ in \cite{dlz2},
and including $\mathbf{120}$ has been done in \cite{lz}.
However, in \cite{dlz2,lz} the MSSM doublets are given by fine-tuning the doublet mass matrix.

The successful doublet-triplet splitting through the DW mechanism in the renormalizable models
has been  realized in \cite{cz},
where the first  $\mathbf{10}$ which couples to matter fields does not couple directly
to a second $\mathbf{10}$ through $\mathbf{45}$.
Instead, a filter sector is proposed since any singlet does not couple
$\mathbf{10}$ to $\mathbf{120}$.
The relevant superpotential is
\be
P H_1 \overline{h}+ m_hh \overline{h}+A h H_2+\frac{1}{2}M_{H_2}^2H_2^2,\label{filter}
\ee
where $P$ is a singlet, $\overline{h}$ and $h$ are all $\mathbf{10}$s and $A$ is a $\mathbf{45}$
whose VEV is in the (15,1,1) direction only.
Following a previous observation\cite{lee} that in a model with  $\mathbf{10,\overline{126},{126},210}$,
if one takes the superpotential as
\be
\Phi \left(H_1+\overline{\Delta}\right)\Delta+M_\Delta\overline{\Delta}\Delta\label{masslesstoy}
\ee
plus terms containing only $\Phi$,
the mass matrix for the doublets is
\be
\left(\begin{array}{c|c}
0_{2\times 3}& A_{2\times 1}\\\hline
B_{2\times 3}& 0_{2\times 1}
\end{array}\right)
\ee
for the bases as ($H_1^u, \overline{\Delta}^u,\Phi^u,{\Delta}^u$) in the columns and
($H_1^d, \overline{\Delta}^d,\Phi^d,{\Delta}^d$) in the rows.
The first three columns are not independent with a combination of them gives a massless $H_u$,
while the first two rows make a massless $H_d$.
The key point is the absence of $H_1\Phi \overline{\Delta}$ which, if present, gives a nonzero
value to the (1,3) matrix element.
Similarly there are also a pair of massless triplets.
After applying the DW mechanism, the triplets become massive and the doublets remain massless.
$\mathbf{120}$ can be included if a pair of them are used,
and the filter sector (\ref{filter}) is needed\cite{cz}.
In \cite{cz} a large $P$ is needed to suppress proton decay.
However, this  makes the main components of
the MSSM doublets are not from $H_1$, so that it is difficult to give the top quark a big mass.

\section{The model}
In the present work we will firstly give a pair of massless doublets and a pair of massless triplets
analogue to the mechanism in \cite{lee,cz}, then use the DW mechanism to give the triples masses.
Since $\mathbf{120}$ is present, a filter sector is needed.
In addition,
we will use the CDW mechanism to forbid  proton decay mediated by $\mathbf{10}$,
so that no large VEV for the singlet $P$ in the filter sector is needed.
The other proton decay amplitudes mediated by $\mathbf{120,\overline{126}}$ are suppressed
by building up the triplet mass matrix analogue to (\ref{masstoy}) but extended.

In order to suppress proton decay mediated by the $\mathbf{120}$ and
$\mathbf{\overline{126} (126})$, we need to double them as $D_{1,2}$ and
$\overline{\Delta}_{1,2},{\Delta}_{1,2}$, respectively.
Besides $H_1$, only $D_1$ and $\overline{\Delta}_{1}$ are allowed to couple with the MSSM matter superfields.
The superpotential for this sector is
\bea
W_{D\Delta}&=&
 \left(k\Phi +m_{\Delta_{12}} \right) \overline{\Delta_1}\Delta_2
+\left(\Phi +m_{\Delta_{21}} \right) \overline{\Delta_2}\Delta_1+Q\Delta_2\overline{\Delta }_2\nonumber\\
&+&\Phi D_1 \left(\Delta_2+\overline{\Delta_2} \right) +\Phi \left(\Delta_1+\overline{\Delta_1} \right) D_2\label{WDD}\\
&+&\Phi H_1\left(D_2+\Delta_2+\overline{\Delta_2} \right)+ \left(m_D+\Phi\right) D_1 D_2+Q D_2^2.\nonumber
\eea
The D- and F-flatness conditions for $v_{(1,2)R},\overline{v}_{(1,2)}$ are the same as
(\ref{dterm},\ref{fvr}) with the seesaw  VEV for $\overline{v_{1R}}$
in (\ref{v1r}) if $Q\sim 10^{-2}\Lambda_{GUT}$.

At first sight, the simultaneous existence of both
$\Phi (H_1+D_1+\overline{\Delta_1}) \Delta_2$ and $\Phi (H_1+D_1+{\Delta_1}) \overline{\Delta_2}$
might invalidate the observation following (\ref{masslesstoy})
that the absence of one of these couplings is the key point of generating massless doublet  and triplet pairs.
The subtlety is that $v_{2R}=0$ given in (\ref{v2r}) as a consequence of  SUSY,
which eliminates the crossing entries between $\Phi$ and $H_1+D_1+\overline{\Delta_1}+{\Delta_1}$ if they
are proportional to $v_{2R}$ in the mass matrices for  the doublets and the triplets,
while those crossing entries exist if they are proportional to $\overline{v_{2R}}$.
The mass matrix for the doublets is
\begin{eqnarray}\label{massDf}
M_D^{D\Delta}=\left(
\begin{array}{c|c}
0_{6\times 5}&A_{6\times 5}\\
\hline
B_{4\times 5}&C_{4\times 5}
\end{array}
\right),
\end{eqnarray}
where the columns are
($H_1^u,D_1^u,D_1^{\prime u},\overline{\Delta}^u_1,{\Delta}^u_1;
{\Phi}^u;\overline{\Delta}^u_2,{\Delta}^u_2 ,D_2^u,D_2^{\prime u}$),
while the rows are similar.
In (\ref{massDf}), the 6th row correponds to $\Phi^d$, and the first 5 entries in this row are
proportional to $v_{2R}$ which is zero, according to (\ref{v2r}).
Then the massless eigenstates are
\bea
H_u^0=\!\!\!\!\sum_{X=H_1,D_1,D_1,\overline{\Delta}_1,{\Delta}_1}\!\!\!\!\!\! \alpha_{X}^u X^u, ~~
H_d^0=\!\!\!\!\sum_{X=H_1,D_1,D_1,\overline{\Delta}_1,{\Delta}_1,\Phi}\!\!\!\!\!\! \alpha_{X}^d X^d.\label{h0}
\eea
Note that the absence of $B-L$ violating component $\Phi^u$ in
$H_u^0$ suggests that there is no   type-II seesaw contribution to the neutrino masses,
as was discussed in \cite{li}.
The relation of large atmospheric mixing and the small quark 2-3
mixing based on the type-II seesaw\cite{bert2,goh} is absence in the present model.

Comparing to the doublets, there are two more pairs of triplets from $\overline{\Delta_{1,2}}+\Delta_{1,2}$.
The mass matrix for the triplets is
\begin{eqnarray}\label{massTf}
M_T^{D\Delta}=\left(
\begin{array}{c|c}
0_{7\times 6}&A_{7\times 6}\\
\hline
B_{5\times 6}&C_{5\times 6}
\end{array}
\right),
\end{eqnarray}
where the columns are\\
($H_1^T,D_1^T,D_1^{\prime T},\overline{\Delta}^T_1,\overline{\Delta}^{\prime T}_1,{\Delta}^T_1;
{\Phi}^T;\overline{\Delta}^T_2,\overline{\Delta}^{\prime T}_2,{\Delta}^T_2 ,D_2^T,D_2^{\prime T}$),
while the rows are similar.
Again, there is a pair of massless triplets.
We can re-write the mass matrix in (\ref{massTf}) as
\begin{eqnarray}\label{massTf2}
M_T^{D\Delta}=\left(
\begin{array}{c|c}
0_{6\times 6}&A^\prime_{6\times 6}\\
\hline
B^\prime_{6\times 6}&C^\prime_{6\times 6}
\end{array}
\right).
\end{eqnarray}
Note that in the lower right sub-matrix $C^\prime_{6\times 6}$,
the (1,1) entry is the mass of the triplet from $\Phi$ which is $\sim\Lambda_{GUT}$,
while the other entries in the first row
are all proportional to $v_{1R}\sim \Lambda_{GUT}$ except one zero.
All the other  entries in the lower 5 rows of $C^\prime_{6\times 6}$ are, besides the zeros,
either proportional to $\overline{v_{1R}}$ or $Q$ which are of $10^{-2}\Lambda_{GUT}$.
Consequently,
there are 5 small eigenvalues in $C^\prime_{6\times 6}$ which are not enough to
generate 6 large effective triplet masses.
The 6th large effective triplet mass will be generated after the application of the CDW mechanism.

To make the massless triplet pair massive while keeping the massless doublets,
we need to apply the DW mechanism.
In realizing both the DW and the CDW mechanisms,
we find three $\mathbf{45}$s ($A,A^\prime,A^{\prime\prime}$) and one $\mathbf{54}$ ($E$)
are needed with the superpotential
\be
W_{DW}=P A A^\prime+(M_{A^\prime A^{\prime\prime}}+E) A^{\prime} A^{\prime\prime}\label{CDW}
\ee
containing all possible interactions of $A^{\prime}$ and $A^{\prime\prime}$ with fields
which may have large VEVs.
As will be seen later, the singlet $P$ can be chosen the same as that in the filter sector.
Labelling $A_1$ and $A_2$ {\it etc.} as the VEVs
in the (1,1,3) and (15,1,1) directions, respectively,
the F-flatness conditions for the VEVs of $A^{\prime\prime}$ are
\bea
0&=&(M_{A^\prime A^{\prime\prime}}+\sqrt{\frac{3}{20}}E)A^\prime_1,\label{f1}\\
0&=&(M_{A^\prime A^{\prime\prime}}-\sqrt{\frac{1}{15}}E)A^\prime_2,\label{f2}
\eea
then either $A^\prime_1$ or $A^\prime_2$ is zero.
When we take the CDW solution $A^\prime_2=0$, (\ref{f1}) gives $M_{A^\prime A^{\prime\prime}}+\sqrt{\frac{3}{20}}E=0$
so that the F-flatness condition for $A_1^{\prime}$ is
\be
0= PA_1,\label{f3}
\ee
(\ref{f3}) is followed by either $P$ or $A_1$ is zero, and we choose the later which is the DW solution.

Accordingly, we will introduce a filter sector which include the singlet $P$ and a pair of
$\mathbf{10}$s ($\overline{h},h$) and use the following superpotential
\be
W_{filter}=
P H_1 \overline{h}+ m_h h \overline{h}+A h H_2+A' H_2 H_3 +\frac{1}{2}M_{H_3}H_3^2.\label{filt}
\ee
Then the relevant mass matrix for the doublets is
\begin{eqnarray}\label{filterMD}
M_D^{filter}=\left(
\begin{array}{ccccc}
0 & \alpha^d_{H_1} P & 0 & 0 &0\\
\alpha^u_{H_1} P & 0 &m_h& 0 &0\\
0 &m_h& 0 &A_1=0&0\\
0 & 0 &A_1=0&0&A_1^\prime\\
0&0&0&A_1^\prime&M_{H_3},
\end{array}
\right)
\end{eqnarray}
where the bases are  $(H_{u(d)}^0, \overline{h}, h, H_2,H_3$)
with $H_{u,d}^0$ given in (\ref{h0}).
Then we have a pair of massless eigenstates
\begin{eqnarray}\label{doub2}
H_u&=&\frac{m_h H_u^0-(\alpha_{H_1}^u P) h^u}{\sqrt{|\alpha_{H_1}^u P|^2+|m_h|^2}}, \nonumber\\
H_d&=&\frac{m_h H_d^0-(\alpha_{H_1}^d P) h^d}{\sqrt{|\alpha_{H_1}^d P|^2+|m_h|^2}},
\end{eqnarray}
which  are the weak doublets in the MSSM.
For $P$ having a VEV of order $\Lambda_{GUT}$,
the components of $H_1^{u,d}$ in the MSSM doublet $H_{u,d}$ are not small,
so that there is no difficulty
in giving the top quark a large mass,
and a large $\tan\beta$ is needed for the small bottom and $\tau$ masses.
Different from the prediction of
$\tan \beta\sim 1$ got by fine-tuning the doublet mass matrix to have light eigenstates\cite{dlz2},
generating these eigenstates through the DW mechanism exhibits the non-trivial aspect of the present work.

Different from the treatment on the weak doublets aiming at giving
predictions on the MSSM doublets,
in the color triplet sector the effective triplet masses which determine proton lifetime are
our main concern.
The triplet mass matrix following $W_{filter}$ is
\begin{eqnarray}\label{filterMT}
M_T^{filter}=\left(
\begin{array}{ccccc}
0 & P & 0 & 0 &0\\
P & 0 &m_h & 0 &0\\
0 &m_h& 0 &A_2&0\\
0 & 0 &A_2&0&A_2^\prime=0\\
0&0&0&A_2^\prime=0 &M_{H_3}
\end{array}
\right),
\end{eqnarray}
where the bases are similar to the doublets.
The effective triplet masses for $H_1$ are generated which are infinity
so that proton decay mediated by $H_1$ is strictly forbidden.
Together with those 5 large effective triplet masses following (\ref{massTf2}),
all the proton decay amplitudes are suppressed.

Alternatively, integrating out only the triplets in $h,H_2,H_3$ while keeping those in $\overline{h}$,
the triplet mass matrix in (\ref{massTf2}) is enlarged by a 13th column
which corresponds to the triplet in $\overline{h}$ and has only one nonzero entry $P$ on the top,
and by a 13th row
which corresponds to the anti-triplet in $\overline{h}$ and has only one nonzero entry $P$ in the left.
Consequently,
the lower-right sub-matrix $C^\prime_{6\times 6}$ in (\ref{massTf2}) is replaced by
a new sub-matrix of ${7\times 7}$
which has 6 small eigenvalues including a zero.
Then integrating out this sub-matrix induces 6 large eigenvalues including an infinity which are
sufficient to suppress all proton decay amplitudes.

Following (\ref{filterMD},\ref{filterMT})
the determinant of $M_T^{filter}$ is $\Lambda_{GUT}^5$,
which is of the same order as
\be
\Lambda_{GUT}{\rm Det}^\prime(M_D^{filter}),\label{thres}
\ee
where ${\rm Det}^\prime(M_D^{filter})=
\lim_{\epsilon\rightarrow 0} \frac{1}{\epsilon}{\rm Det}(M_D^{filter}+\epsilon {\rm\hat I}_{4\times 4})$
is the product  of all  masses of the doublets except the MSSM doublets\cite{msso10b,msso10c},
so there is no large threshold effect in the doublet-triplet sector relevant for gauge coupling unification.
Possible threshold effects from other states can be included following \cite{al1,al2,al3,al4},
so that gauge coupling unification can be fully realized by adjusting the parameters of the model.

The above results need to be protected by an extra symmetry.
After very difficult efforts,
we find that the $Z_{24}\times Z_4$ symmetry can be used.
In addition,
to maintaining SUSY at $\Lambda_{GUT}$,
we need to introduce  one singlet ($R$)
replacing $M_{A^\prime A^{\prime\prime}}$ in (\ref{CDW}) and a second
$\mathbf{54}$ ($E^{\prime}$).
Also, a new $\mathbf{45}$ ($A'''$) is introduced to kill redundant massless states whose existences break unification.
Under the $Z_{24}\times Z_4$ symmetry,
the transformation properties of all the particles are listed in Table 1.

\begin{table}\begin{center}
\begin{tabular}{|c||c|c|c|c|c|c|c|c|c|c|c|}
  \hline  \hline
   & $A^{\prime\prime\prime}$ &$A^{\prime\prime}$ & $E$ & $R$ & $A^{\prime}$ & $P$ & $A$ & $E^{\prime}$ &$\Phi$& $Q$ & $\psi_i$\\\hline
  $Z_{24}$ & 12&2 & 12 & 12 & 10 & 2 & 12 & 0 & 0 &4&-1\\\hline
  $Z_4$&0&-1&0&0&1&1&2&0&0&0&0\\\hline\hline
      & $H_1$& $\overline{h}$ & $h$& $H_2$&$H_3$ & $D_1$ &$\overline{\Delta}_1$ & ${\Delta}_1$ &$D_2$ &$\overline{\Delta}_2$ & ${\Delta}_2$ \\
  \hline
  $Z_{24}$ & 2&-4&4&8&6&2&2&2&-2&-2&-2\\\hline
    $Z_{4}$ &0&-1&1&1&2&0&0&0&0&0&0\\
  \hline  \hline
\end{tabular}
\caption{$Z_{24}\times Z_4$ properties of all superfields. Here $\psi_i (i=1,2,3)$ are matter superfields.}\end{center}
\end{table}

The complete superpotential is
\be
W^{Full}=W_{DW}^{Full}+W_{SB}^{Full}+W^{Full}_{D\Delta}+W_{filter}^{Full},
\ee
where
\bea
W_{DW}^{Full}&=&P A A'+(E+R) A' A''+A'A''A''',\nonumber\\
W_{SB}^{Full}&=&\frac{1}{2}m_{\Phi }\Phi ^2 +\Phi ^3+E^\prime \Phi^2+\Phi A^2+\frac{1}{2}m_{{E'}} {E'}^2
+\frac{1}{2}m_EE^2\nonumber\\
&+&E^2 {E'}+{E'}^3+\frac{1}{2}m_AA^2+{E'}A^2 +\frac{1}{2} m_R R^2+R E E^\prime\nonumber\\
&+&\frac{1}{2}m_{A'''}A^{\prime\prime\prime 2}+E^\prime A^{\prime\prime\prime 2}+\Phi A^{\prime\prime\prime 2},\nonumber\\
W_{D\Delta}^{Full}&=&
 \left(k\Phi +m_{\Delta_{12}} \right) \overline{\Delta_1}\Delta_2
+\left(\Phi +m_{\Delta_{21}} \right) \overline{\Delta_2}\Delta_1+Q\Delta_2\overline{\Delta }_2\nonumber\\
&+&\Phi D_1 \left(\Delta_2+\overline{\Delta_2} \right) +\Phi \left(\Delta_1+\overline{\Delta_1} \right) D_2
+Q D_2^2\nonumber\\
&+&E^\prime (\Delta_1\Delta_2+\overline{\Delta_1}\overline{\Delta_2})\nonumber\\
&+&\Phi H_1\left(D_2+\Delta_2+\overline{\Delta_2} \right)
+\left(m_D+{E'}+\Phi\right) D_1 D_2,\nonumber\\
W_{filter}^{Full}&=&P H_1 \overline{h}+ \left({E'}+m_h\right)h \overline{h}+A h H_2+A' H_2 H_3 \nonumber\\
&+&\frac{1}{2}(R+E)H_3^2.\nonumber
\eea
Obviously,
$W_{DW}^{Full}$ gives the DW and CDW solutions for $A$ and $A^\prime$, respectively.
Its last term does not contribute to any of the F-flatness conditions.
$W_{SB}^{Full}$ is the main sector breaking SO(10).
$W_{D\Delta}^{Full}$, whose first terms  break $U(1)_{I_{3R}}\times U(1)_{B-L}$ into $U(1)_Y$,
generates a pair of massless doublets and a pair of massless triplets.
The first three terms in $W_{filter}^{Full}$ generate  a  coupling $H_1 (P) A H_2$ at the same time forbidding
the coupling between $H_2$ and $D_{1}$ through $A$,
so that in both $H_{1,2}$ the doublets are massless and the triplets are massive.
The last two terms in $W_{filter}^{Full}$ gives masses to the doublets in $H_2$.
Accordingly there are some modifications without any important changes in the results.

We have examined all the F-flatness conditions for the SM singlets without finding any conflict.
There are two subtleties in these conditions.
For the singlet VEV $P$,  the condition is
\be    0=A_1 A'_1+A_2 A'_2, \nonumber\ee
which is automatic following $A_1=0$ and  $A'_2=0$, the DW and the CDW solutions, respectively.
This is due to the existence of an accident Peccei-Quinn like U(1) symmetry\cite{PQS}.
Consequently, in other flatness-conditions,
the VEV of $P$ only appears in products with other VEVs of the GUT breaking fields,
so that it is natural to take the VEV of $P$ the same order as $\Lambda_{GUT}$,
which makes $P$ a harmless axion\cite{axion}.

The second subtlety is that for the singlet $Q$,
the F-flatness condition
\be
0=\overline{v_{2R}}  {v}_{2R},\nonumber
\ee
which  is also automatic since ${v}_{2R}=0$.
$Q$ does not enter any other condition for keeping SUSY except (\ref{fvr}),
so that it can be given a value of the seesaw scale VEV $\Lambda_{seesaw}\sim 10^{-2}\Lambda_{GUT}$,
or be generated to be $\frac{\Lambda_{GUT}^2}{M_{Planck}}$ of the same order\cite{lz}
through the Green-Schwarz mechanism\cite{green1,green2,green3,green4}.

\section{Summary}
We have  proposed in the present work  a renormalizable SUSY SO(10)
with the following results.
First,
we use  a seesaw  VEV instead a seesaw scale so that
SO(10) breaks directly into the SM gauge group without spoiling gauge coupling unification.
We find  an important point ($v_{2R}=0$) in generating massless states before apply the DW mechanism.
Second,
naturally doublet-triplet splitting is realized through the DW mechanism using a filter sector,
and the DW and CDW mechanisms are very simply realized.
Third,
proton decay is suppressed successfully through the realization of CDW mechanism and through
the special structure of the color-triplet mass matrix.
Especially, the proton decay amplitude mediated by $\mathbf{10}$, which couples with the MSSM matter superfields
with the largest coupling,  is strictly forbidden.

Although the present model is complicated, it has solved the main difficulties
and is thus the first realistic SUSY GUT model.
Large representations used in in the present model,
as well as in all other renormalizable SUSY GUT models,
bring in the result that the GUT gauge coupling blows up quickly above the GUT scale.
This result can be explained if  we take the picture that in the very early universe
there was a phase transition of the GUT symmetry breaking.
Consequently,
without a very clear understanding on the details during this phase transition,
the non-perturbative behavior of the GUT gauge coupling above the GUT scale  may not be
a real problem.

We thank Xiaojia Li for many discussions.

\end{document}